\def\kF{k_{\text{F}}}
\def\TF{T_{\text{F}}}
\def\Tc{T_{\text{c}}}
\def\Ttc{T_{\text{tc}}}
\def\vF{v_{\text{F}}}

\def\ne{n_{\text{e}}}

\def\muB{\mu_{\text{B}}}
\def\kB{k_{\text{B}}}
\def\gammat{\gamma_{\text{t}}}
\def\muOhmcm{\mu\Omega{\rm cm}}

\def\be{\begin{equation}}
\def\ee{\end{equation}}
\def\bea{\begin{eqnarray}}
\def\eea{\end{eqnarray}}
\def\bse{\begin{subequations}}
\def\ese{\end{subequations}}

\documentclass[prl,twocolumn,showpacs,amsmath,amssymb,preprintnumbers]{revtex4}
\usepackage{graphicx}
\usepackage{dcolumn}
\usepackage{bm}
\begin{document}
\title{Disorder dependence of the ferromagnetic quantum phase transition}
\author{Y. Sang$^1$, D. Belitz$^{1,2}$ and T.R. Kirkpatrick$^3$}
\affiliation{$^{1}$ Department of Physics and Institute of Theoretical Science,
                    University of Oregon, Eugene, OR 97403, USA\\
                   $^{2}$ Materials Science Institute, University of Oregon, Eugene,
                    OR 97403, USA\\
                  $^{3}$ Institute for Physical Science and Technology,
                    and Department of Physics, University of Maryland, College Park,
                    MD 20742, USA
            }
\date{\today}

\begin{abstract}
We quantitatively discuss the influence of quenched disorder on the ferromagnetic 
quantum phase transition in metals, using a theory that describes the coupling 
of the magnetization to gapless fermionic excitations. In clean systems, the transition 
is first order below a tricritical temperature $\Ttc$. Quenched disorder is predicted to 
suppress $\Ttc$ until it vanishes for residual resistivities $\rho_0$ on the order of several  
$\muOhmcm$ for typical quantum ferromagnets. We discuss experiments that 
allow to distinguish the mechanism considered from other possible realizations of a 
first-order transition.
\end{abstract}


\maketitle

\begin{figure}[t]
\vskip -0mm
\includegraphics[width=8.0cm]{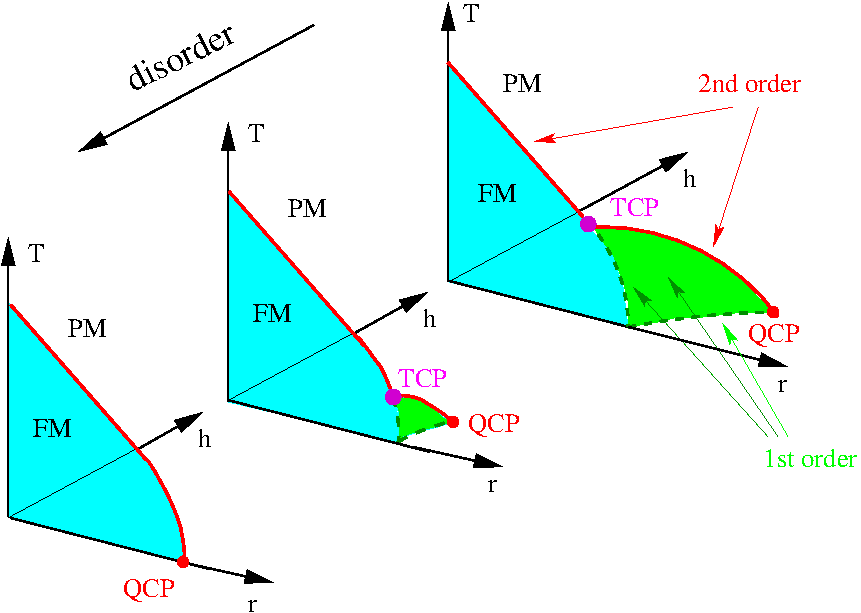}
\caption{(Color online) Evolution of the phase diagram of a metallic quantum ferromagnet in the 
               space spanned by temperature ($T$), magnetic field ($h$), and the 
               control parameter ($r$) with increasing disorder. Shown are the ferromagnetic (FM) and 
               paramagnetic (PM) phases, lines of second-order transitions (solid red), the
               tricritical point (TCP), and a surface of first-order transitions  (``tricritical 
               wing'') that ends in a quantum critical point (QCP). 
               With increasing
               disorder the tricritical temperature decreases, the wings shrink, and
               above a critical disorder strength a QCP is realized in zero field. 
               }
               \vskip -0mm
\label{fig:1}
\end{figure}

There is substantial experimental evidence for the quantum ferromagnetic transition
in clean metals to be generically of first order. Examples of systems that were
expected to display a quantum critical point, but instead display a first-order transition
if the Curie temperature is suppressed, are
MnSi \cite{Pfleiderer_et_al_1997, Pfleiderer_Julian_Lonzarich_2001}, 
ZrZn$_2$ \cite{Uhlarz_Pfleiderer_Hayden_2004}, 
UGe$_2$ \cite{Huxley_et_al_2001, Taufour_et_al_2010, Kotegawa_et_al_2011}, and 
URhGe \cite{Huxley_et_al_2007}. All of these are low-temperature ferromagnets
(although the magnetic moment in some of them is not small) with Curie temperatures $\Tc$ 
between $\approx 10$K (URhGe) and $\approx 50$K (UGe$_2$), and magnetic 
moments per formula unit of about $0.17$, $0.4$, and $1.5\,\muB$ for ZrZn$_2$,
MnSi and URhGe, and UGe$_2$, respectively. $\Tc$ is tunable by hydrostatic pressure or, for URhGe,
by an external magnetic field transverse to the easy axis.  A
tricritical point separates a line of second-order
transitions above the tricritical temperature $\Ttc$ ($\approx 5$K, $10$K, $1$K, and $24$K
in ZrZn$_2$, MnSi, URhGe, and UGe$_2$, respectively) from a line of first-order transitions
below, and in all of these materials tricritical wings have been observed in an
external magnetic field. The respective values of the critical field at the wing tips are
$H_{\text{c}} \agt 0.05$T, $\approx 0.6$T, $\approx 1$T, and $\approx 10$T. The qualitative phase 
diagram is shown in the rightmost panel in Fig.\ \ref{fig:1}. 
Evidence for a first-order quantum phase transition (QPT) at low temperatures
has been found in many other systems, but the phase diagram has not been
mapped out completely, or the tricritical point is not accessible
(as in UCoAl \cite{Aoki_et_al_2011b}). 

These observations are remarkable because of their universality. The only
known instances in which the QPT to a homogeneous ferromagnet is not observed to be of first order are the
quasi-1-d material YbNi$_4$P$_2$ \cite{Krellner_et_al_2011}, where the physics is expected to be quite
different from that of true bulk metals, and various
disordered materials where the transition is tuned by chemical
composition, e.g., URu$_{2-x}$Re$_x$Si$_2$ \cite{Butch_Maple_2009}. 
The weakly disordered compounds
Ni$_x$Pd$_{1-x}$ \cite{Nicklas_et_al_1999} and (Cr$_{1-x}$Fe$_x$)$_2$B \cite{Schoop_et_al_2014}
we will come back to.
Also remarkable is the stark disagreement between experiment
and early theories. The quantum ferromagnetic transition as described
by Stoner's mean-field theory \cite{Stoner_1938} is generically continuous.
It was later considered as an example
by Hertz in his seminal renormalization-group treatment of
QPTs \cite{Hertz_1976}, which also predicted a continuous
transition with mean-field critical behavior.

A theory that explains, and indeed predicted, the observed universality was
developed in Refs.\ \onlinecite{Belitz_Kirkpatrick_Vojta_1999, Belitz_Kirkpatrick_Rollbuehler_2005, 
Kirkpatrick_Belitz_2012b}. It relies on the coupling between the magnetization and soft or gapless 
fermionic excitations with a ballistic frequency-momentum relation that exist in any clean metal at 
$T=0$. It leads to an equation of state of the form
\be
h = r\,m - v\,m^3\ln (1/m^2) + u\,m^3\ ,
\label{eq:1}
\ee
where $m$ is the magnetization in suitable units, $h$ is the external field, $r$ is
the control parameter, and $v$ and $u$ are Landau parameters. The nonanalytic
$m^3\ln(1/m)$ term is the result of $m$ coupling to the soft modes in $d=3$; more 
generally its form is $m^d$. Crucially, $v>0$, which leads to a first-order transition at some $r>0$. 
This {\em universal} mechanism has been confirmed by a variety of other techniques 
\cite{Chubukov_Pepin_Rech_2004, Karahasanovic_Kruger_Green_2012, homogeneous_FM_footnote}.
$T>0$ gives the soft modes a mass, which cuts off the $\ln m$ nonanalyticity and leads to
a tricritical point (rightmost panel in Fig.\ \ref{fig:1}). 
Quenched disorder has two effects. First, it also cuts off the $\ln m$. Second, 
a coupling to diffusive soft modes leads to an $m^{d/2}$ nonanalyticity whose sign is opposite of that of the
nonanalytic term in the clean case. For sufficiently strong disorder,
in $d=3$, one finds \cite{Kirkpatrick_Belitz_1996}
\be
h = r\,m + \frac{w}{(\kF\ell)^{3/2}}\,m^{3/2} + u\,m^3\ ,
\label{eq:2}
\ee
with $w>0$ another parameter, $\kF$ the Fermi wave number, which sets the microscopic
length scale, and $\ell$ the elastic mean-free path. This leads to a continuous transition with
non-mean-field exponents (leftmost panel in Fig.\ \ref{fig:1}). Equations (\ref{eq:1}) and (\ref{eq:2}) both represent
renormalized Landau theories, which replace the fluctuating order-parameter field
by its mean. Since the order-parameter fluctuations at the QPT
are above their upper critical dimension ($d_{\text{c}}^+ = 1$ ($0$) for the
clean (disordered) case \cite{Hertz_1976}), this should not affect
the nature of the QPT. Indeed, in the disordered case a RG study has shown 
that order-parameter fluctuations leave the power laws described by Eq.\ (\ref{eq:2}) intact,
although they lead to non-power-law modifications of the leading scaling 
behavior \cite{Belitz_et_al_2001a, Belitz_et_al_2001b}.

While Eq.\ (\ref{eq:1}) is in qualitative agreement with all experiments on clean samples,
there still are open questions. First, Eqs.\ \ (\ref{eq:1}) and (\ref{eq:2}) represent
only the extremes of ultraclean and strongly disordered systems. Many experiments
fall in between these two cases, e.g., Ni$_x$Pd$_{1-x}$, which shows a
second-order transition with mean-field exponents at
$x = 0.027$ \cite{Nicklas_et_al_1999}. Second, the mechanism proposed in 
Ref.\ \onlinecite{Belitz_Kirkpatrick_Vojta_1999} is not the only possibility for a first-order
transition; e.g., in classical compressible magnets the coupling between the
magnetization and phonons can lead to a first-order transition \cite{Aharony_1976}.
This mechanism is not as universal as the one leading to Eq.\ (\ref{eq:1}), but
Refs.\ \onlinecite{Gehring_2008, Gehring_Ahmed_2010, Mineev_2011} have argued
that an adaptation to the quantum transition can explain the
observations, at least in the case of the pressure-tuned quantum ferromagnets. 

It is thus desirable to develop criteria that allow for a
discrimination between the different theoretical ideas. In this Letter we
show that the disorder dependence of the phase diagram allows for such a discrimination.
Phonons are not qualitatively affected by disorder; therefore, if magnetostriction effects
cause the first-order transition in a given material, then introducing disorder into
the sample is expected to have only weak quantitative effects on the phase diagram.
The mechanism of Ref.\ \onlinecite{Belitz_Kirkpatrick_Vojta_1999}, on the other hand, is
crucially affected by disorder, since in the strong-disorder limit the equation of state changes
to Eq.\ (\ref{eq:2}). 
As we will show, our theory predicts three distinct disorder regimes.
In a weak-disorder regime the transition is first
order, but $\Ttc$ is gradually suppressed until it vanishes at a critical
value of the disorder. For common quantum ferromagnets
this is expected to happen for residual resistivities $\rho_0$ on the order of several $\mu\Omega$cm. 
The resulting quantum critical point in an intermediate-disorder regime displays mean-field exponents 
consistent with Hertz theory in the observable critical region, although asymptotically
close to the transition there will be a crossover to the non-mean-field critical
behavior of Ref.\ \onlinecite{Kirkpatrick_Belitz_1996}. With increasing disorder the crossover
moves away from the transition and becomes observable for values of
$\rho_0$ on the order of tens of $\mu\Omega$cm. Finally, in a strong-disorder regime with $\rho_0$ on the
order of hundreds of $\mu\Omega$cm the non-mean-field critical behavior 
will be present in the entire critical region. However, for disorder that
strong other effects may come into play. Our predictions can be tested
by introducing quenched disorder, e.g., by means of irradiation, into any of the
materials that display a first-order QPT, and following the
changes in the phase diagram with increasing disorder.

To achieve these goals, we have constructed an equation of state that interpolates between
Eqs.\ (\ref{eq:1}) and (\ref{eq:2}), and generalizes them to finite temperatures. We first state 
and discuss this equation of state, and then sketch its derivation. It takes the form
\bse
\label{eqs:3}
\bea
h &=& r\,m + \frac{w}{(\kF\ell)^{3/2}}\,m^{3/2}\,g(\kF\ell\,m, c\,t/m)
\nonumber\\
             &&\hskip -20pt  - v\, m^3 \ln \left(\frac{1}{m^2/m_0^2 + (\sigma_0/\kF\ell + t)^2}\right) + u\,m^3\ ,\quad
\label{eq:3a}
\eea
which reduces to Eqs.\ \ (\ref{eq:1}), (\ref{eq:2}) in the limits $\kF\ell \to \infty$, $0$.
We will refer to the second and third term on the right-hand side as the diffusive and ballistic
nonanalyticity, respectively.
$m$, $h$, and $t$ are the dimensionless magnetization, magnetic field, and temperature,
respectively, defined as follows. Let $\mu$ be the magnetization measured in $\muB$ per volume,
$H$ the external field, and $T$ the temperature. Let $\ne$ be the conduction electron
density, and $\TF$ the Fermi temperature (or, more generally, the microscopic energy scale).
Then $m = 8\mu/\pi\ne$, $h = \muB H/\kB\TF$, and $t = 3\pi T/\TF$. $v$ and $w$ depend on a coupling constant
${\gamma}_{\text{t}}$ that measures the strength of conduction-electron correlations,
with $\gammat\ll 1$ and $\gammat=O(1)$ corresponding to weakly and strongly
correlated systems, respectively. Another coupling constant $c = O(1)$ describes the coupling between the
magnetization and the conduction-electron spin density. In terms of 
$\gammat$ and $c$ one finds, for small $\gammat$,
$v = c\,{\gamma}_{\text{t}}^4$ and $w = c\,{\gamma}_{\text{t}}$. $r\ll 1$ is the
dimensionless control parameter for the transition, and $u = O(1)$ is a Landau parameter. 
$m_0$ and $\sigma_0$ set the scales for the magnetic moment
and the disorder, respectively.
In a simple model for itinerant ferromagnets one has $m_0 \approx 7/\gammat$
and $\sigma_0 = 1$; more generally $m_0$ and $0.1 \alt \sigma_0 \leq 1$ are independent 
microscopic scales that depend on the band structure and the correlation strength, see the discussion below.
Finally, 
\be
g(y,z) = \frac{1}{g_0} \int_0^{1/y} \hskip -10pt dx  \int_z^{\infty} \hskip -8pt d\omega\,
   \frac{\sqrt{x}\,\omega[2(x+\omega)^2 + 1]}{(x+\omega)^3 [(x+\omega)^2+1]^2}
\label{eq:3b}
\ee
with $g_0 = \pi/3\sqrt{2} \approx 0.74$ is normalized such that $g(0,0) = 1$. $g(y,0)$ is well approximated by
\be
g(y,0) \approx 1/[1+y^{3/2}/(9g_0 + y/g_0)]\ .
\label{eq:3c}
\ee
\ese

We now discuss typical values for the various parameters in Eq.\ (\ref{eq:3a}), initially for a clean system.
With $\kF \approx 1\,\AA^{-1}$, and a formula unit volume of about $50\,\AA^3$, we find
a dimensionless saturation magnetization ranging from $m \approx 0.25$ for ZrZn$_2$ to  $m \approx 2.3$ for UGe$_2$. 
Choosing $u=0.85$, $\gammat = 0.5$ (fairly strong correlation), and $c=1$, we
have $v = 0.06$. The tricritical temperature is $\Ttc = (\TF/3\pi)  e^{-u/2v}$  
\cite{Belitz_Kirkpatrick_Vojta_1999}. With $\TF \approx 10^5$ K we have $\Ttc \approx 10$ K, 
which is the correct order of magnitude for ZrZn$_2$, MnSi, and UGe$_2$. A slightly lower value of
$\gammat \approx 0.45$ yields $\Ttc \approx 1$ K, as observed in URhGe. 
At the first-order transition at $T=0$, the magnetization changes discontinuously from zero to 
$m_1 = m_0\, e^{-(1+u/v)/2}$ \cite{Belitz_Kirkpatrick_Vojta_1999}. For $m_0$ between 75 
(for ZrZn$_2$) and 350 (for UGe$_2$), this yields
$m_1 \approx 0.05$ - $0.25$, which is a reasonable fraction of the saturation magnetization in these
materials. The critical field at the tips of the tricritical wings is given by 
$h_c = (4/3) e^{-13/4}\,m_0^3\,v\,e^{-3u/2v}$ \cite{Belitz_Kirkpatrick_Rollbuehler_2005}. 
With parameters as above this yields values from
$H_c \approx 0.1$ T to $H_c \approx 10$ T. This is again the correct order of magnitude compared with
the experimental observations \cite{Uhlarz_Pfleiderer_Hayden_2004, Taufour_et_al_2010, Pfleiderer_Julian_Lonzarich_2001}.

Now 
consider quenched disorder. 
A Drude formula for $\rho_0$ with $\kF \approx 1\,\AA^{-1}$ yields $\kF\ell \approx 1,000 \muOhmcm/\rho_0$. 
$\kF\ell$ thus ranges from $\agt 10^4$ in a clean metal ($\rho_0 \approx 0.1\muOhmcm$) to about $10$ in a poor metal 
($\rho_0 \approx 100\muOhmcm$). This in turn implies that values of $\kF\ell\,m$ between roughly $2.5$ and $2\times 10^4$ 
are realizable, with $m$ the saturation magnetization. With $m$ the actual magnetization, the lower limit is 
accordingly lower, depending on the minimal magnetization $m_1$ at the first-order transition, if any.
From Eq.\ (\ref{eq:3c}) we see that $\kF\ell\,m \approx 5$ is the demarcation between two different regimes,
which falls well within this range. 

All of the above, and everything that follows, are just rough
order-of-magnitude estimates. With this in mind, we can distinguish the following regimes, classified according
to the values of $\kF\ell$ (clean vs. dirty samples) and $m$ (weak vs strong magnetism).
They follow from the observation that the diffusive and ballistic nonanalycities, at $T=0$, are operative (inoperative) for
$\kF\ell\,m \alt 5$ ($\agt 5$) and $\kF\ell\,m \agt m_0\,\sigma_0$ ($\kF\ell\,m \alt m_0\,\sigma_0$), respectively.
%
\vskip 2pt
\paragraph{Regime I (Clean/strong):} $\kF\ell\,m \agt m_0\,\sigma_0$. The diffusive nonanalyticity is inoperative,
the equation of state is given by Eq.\ (\ref{eq:1}), and the transition is first order with 
$m_1 = m_0\, e^{-(1+u/v)/2} \leq m$. For consistency, we must have $\kF\ell\,m_1 \agt m_0\,\sigma_0$. With $u$ and $v$ 
as above, and $\sigma_0 \approx 1/5$, this yields $\kF\ell \agt 300$, or $\rho_0 \approx$ several $\muOhmcm$.
\vskip 2pt
\paragraph{Regime IIa (Intermediate):} $5 \alt \kF\ell\,m \alt m_0\,\sigma_0$. In this transient regime both
nonanalyticities are inoperative, and the transition appears continuous with mean-field
exponents in a range of $m$-values. However, as $m$ 
decreases, the system eventually enters Regime IIb or III. 
%
\vskip 2pt
\paragraph{Regime IIb (Intermediate):} $\kF\ell\,m \alt 5$ and $\kF\ell \agt (\kF\ell)^*$, with $(\kF\ell)^*$ defined below. 
The ballistic nonanalyticity is inoperative, the equation of state is given by Eq.\ (\ref{eq:2}), and 
the transition is second order with the asymptotic critical behavior characterized by the non-mean-field exponents 
of Ref.\ \cite{Kirkpatrick_Belitz_1996}. However,
farther away from the transition this behavior will cross over to ordinary mean-field behavior at a disorder-dependent
value $r^*$ of $r$. The crossover occurs when the last two terms in Eq.\ (\ref{eq:2}) are
about equal. Having the crossover occur at $r = r^*$ thus requires a disorder given by
$\kF\ell = \kF\ell^* = w^{2/3}/u^{1/6} \vert r^*\vert^{1/2}$. If we require $r^* = 0.01$ and choose $\gammat = 0.5$
and $u=1$ as before, we have $\kF\ell^* \approx 6$, or $\rho_0^* \approx 150 \muOhmcm$. $\rho_0^*$ is the 
disorder that separates Regime IIb, where the transition is continuous with effectively
mean-field exponents, from Regime III. Note that $\rho_0^*$ depends on the correlation strength via $w$; 
for $\gammat = 0.1$ (weak correlation) one has $\rho_0^* \approx 500 \muOhmcm$.
%
\vskip 2pt
\paragraph{Regime III (Dirty/weak):}  $\kF\ell\,m \alt 5$ and $\kF\ell \alt (\kF\ell)^*$. The equation of state is
dominated by the diffusive nonanalyticity, and the transition is continuous with non-mean-field critical
exponents in the entire critical region. This requires $\rho_0 > \rho_0^*$, with $\rho_0^*$ ranging from approximately
$100 \muOhmcm$ for strongly correlated materials to hundreds of $\muOhmcm$ for weakly correlated ones.

\smallskip
At a nonzero temperature, we see from Eq.\ (\ref{eq:3a}) that a
disorder resulting in $\kF\ell = \sigma_0\TF/3\pi\Ttc$ has the same effect as $T=\Ttc$ in
a clean system. That is, $\rho_0 \agt 10^4\,\Ttc/\sigma_0\TF \approx$ several $\muOhmcm$ will suppress $\Ttc$ to zero, 
consistent with the above estimate for the destruction of the first-order
transition at $T=0$. The tricritical wings shrink, and eventually disappear, commensurate with the 
suppression of $\Ttc$. This prediction for the evolution of the phase diagram is shown 
in Fig.\ \ref{fig:1}.

We now have the following predictions for the effects of quenched disorder on typical strongly correlated quantum ferromagnets:
Disorder decreases $\Ttc$, and suppresses it altogether for a residual resistivity $\rho_0$ on the order of
several $\muOhmcm$. For larger $\rho_0$ the QPT will be continuous and appear mean-field-like
in a substantial disorder range, $\rho_0 \alt 100 \muOhmcm$, with a crossover to non-mean-field
behavior only extremely close to the transition. For even larger $\rho_0$ the critical behavior is characterized by the
non-mean-field exponents of Refs.\cite{Kirkpatrick_Belitz_1996, Belitz_et_al_2001a, Belitz_et_al_2001b}. 
However, for disorder that strong 
quantum Griffiths effects are expected to be present and compete with the critical behavior \cite{Vojta_2010}; to
distinguish between the two one needs to measure the critical behavior of the magnetization. We stress that
these predictions are semi-quantitative in nature. The important point is the existence of the three regimes; the
disorder strengths that delineate them are expected to show substantial variations from material to material.
We also note that quenched disorder can suppress a tricritical point 
in a purely classical model \cite{Falicov_Berker_1996}. This mechanism is not dependent on the presence of 
conduction electrons and is expected to be characterized by different disorder scales than the one discussed here.

Some experimental evidence exists in favor of this scenario. For ZrZn$_2$, the QPT
was initially found to be second order \cite{Grosche_et_al_1995}, but with increasing sample quality
a first-order transition emerged \cite{Uhlarz_Pfleiderer_Hayden_2004}. For UGe$_2$, measured values of $\Ttc$ 
range from $24$K \cite{Taufour_et_al_2010} to $31$K \cite{Huxley_et_al_2007}, which is possibly
related to the sample quality, and in URhGe higher $\Ttc$ values were found for cleaner samples \cite{Huxley_2013}.
All of these materials are strongly correlated as evidenced by their unusual electronic properties independent of the
quantum magnetism. Finally, the observation of a quantum critical point with mean-field exponents in Ni$_x$Pd$_{1-x}$
\cite{Nicklas_et_al_1999},
and possibly in (Cr$_{1-x}$Fe$_x$)$_2$B \cite{Schoop_et_al_2014},
where the transition occurs at a small value of $x$, can be understood if one realizes that these
system are likely in the intermediate Regime II.
While these observations are encouraging, no systematic experimental study of the influence of quenched
disorder on the phase diagram of quantum ferromagnets
exists. Such an experiment would allow to discriminate between the explanation of the first-order
transition discussed above and alternative proposals that predict only a weak disorder dependence of $\Ttc$.

We now sketch the derivation of Eq.\ (\ref{eq:3a}). The relevant soft fermionic modes, 
as functions of wave vector ${\bm k}$ and bosonic Matsubara frequency $\Omega_n$, are diffusive for disordered 
electrons and ballistic for clean ones \cite{Belitz_Kirkpatrick_Vojta_2005},
\bse
\label{eqs:4}
\be
{\cal D}_{\rm{diff}} = \frac{1}{\Omega_n + D{\bm k}^2}\quad,\quad{\cal D}_{\rm{ball}} = \frac{1}{\Omega_n + \vF\vert{\bm k}\vert}
\label{eq:4a}
\ee
\ese
with $D = \vF^2\tau/3$ the diffusion coefficient, $\tau$ the elastic mean-free time, and
$\vF$ the Fermi velocity. The soft-mode propagator ${\cal D}$ can be modeled by 
${\cal D}_{\rm{diff}}$ for $\vert{\bm k}\vert < 1/\ell$, and by ${\cal D}_{\rm{ball}}$ for $\vert{\bm k}\vert > 1/\ell$, with
$\ell = \vF\tau$ the elastic mean-free path. The magnetization $m$ couples to the soft fluctuations and cuts off the
singularities that result from integrating over ${\cal D}$, which leads to nonanalytic dependences on $m$. 
Integrating out the soft modes yields a
fluctuation correction to the free energy density $f$ of the form \cite{Belitz_Kirkpatrick_Vojta_2005}
\be
\Delta f = \frac{2}{V}\sum_{\bm k} T\sum_{n=1}^{\infty} \ln\,N({\bm k},\Omega_n;m)\ .
\label{eq:5}
\ee
At $T=0$ the sum over $\Omega_n$ turns into an integral over a continuous variable $\omega$,
and the effect of a nonzero temperature can be modeled by the replacement $\omega \to \omega + 2\pi T$.
The fluctuation contribution to the equation of state
is obtained by differentiating $\Delta f$ with respect to $m$. We 
measure $m$ in units of the conduction electron
density $\ne$, and the magnetic field $h$ in units of $\kB\TF/\muB$.
The Landau parameters $r$ and $u$ are then dimensionless. Up to
factors of $O(1)$, the resulting equation of state takes the form of Eq.\ (\ref{eq:3a})
with $m_0 \approx 7/\gammat$ and $\sigma_0 = 1$. These two values are based on a nearly-free-electron
model for the conduction electrons. For real materials, $m_0$ is expected to be an independent parameter that
depends on microscopic details. It sets the scale for the magnetic moment, which differs
by a factor of $10$ between, e.g., ZrZn$_2$ and UGe$_2$. $\sigma_0$ in general depends
on the correlation strength and is $\leq 1$. The reason is that in a strongly correlated material  two electrons 
with opposite spins cannot simultaneously take advantage
of a disorder-induced potential well, because of the strong repulsion between
the electrons. This is consistent with the fact that, in the absence of symmetry-breaking fields, interactions cause
the disorder to get renormalized downward \cite{Finkelstein_1983, Belitz_Kirkpatrick_1994}. Correlations will thus
effectively weaken the effects of the disorder; values of $\sigma_0$ between 1 (no correlation) 
and 0.1 (strong correlation) are reasonable based on the RG flow equations of Ref.\ \onlinecite{Finkelstein_1983}. 
Finally, the soft-mode effects are stronger the lower the dimension;
in $d=2$, 
the $m^3 \ln (1/m)$ term in Eq.\ (\ref{eq:3a}) turns into an $m^2$ term. For the diffusive modes, $d=2$ is
the lower critical dimension, and the effects of quenched disorder become strong and very complex \cite{Belitz_Kirkpatrick_1994}.

\acknowledgments
We thank Greg Stewart, Andrew Huxley, and Nihat Berker for discussions. This work was supported by
the NSF under grant Nos. DMR-09-29966 and DMR-09-01907. Part of this work was performed at the
Aspen Center for Physics and supported by the NSF under Grant No. PHYS-1066293.


\end{document}